\documentclass[12pt]{article}
\usepackage{color,amsmath,amssymb,amsthm,bm,epsf,epsfig,enumerate,alltt,latexsym,graphicx}
\usepackage[toc,page,titletoc]{appendix}
\usepackage[T1]{fontenc}
\usepackage{mathabx,textcomp,stmaryrd}
\newcommand{\E}{\ensuremath{\mathrm{E}}}
\newcommand{\var}{\ensuremath{\mathrm{Var}}}
\newcommand{\pr}{\ensuremath{^{\prime}}}
\newcommand{\id}{\ensuremath{\mathrm{I}}}
\renewcommand{\P}{\ensuremath{\mathbf{P}\!}}
\newcommand{\N}{\ensuremath{\mathbf{N}}}
\renewcommand{\sp}{\ensuremath{\mathrm{sp}}}
\newcommand{\tr}{\ensuremath{\mathrm{tr}}}

\newcommand{\Diag}{\ensuremath{\mathrm{Diag}}}
\newcommand{\mbk}{\ensuremath{\mathbb{K}}}
\newtheorem{propn}{Proposition}

\newcommand{\bc}[1]{\ensuremath{\mathbf{\mathcal{#1}}}}
\newcommand{\imp}{\ensuremath{\Longrightarrow}}

\begin{document}
\begin{center}
\begin{Large}
{\bfseries Deconstructing Type III}\end{Large}\\
\vspace{.5cm}by Lynn R. LaMotte\footnote{School of Public Health, LSU Health, New Orleans, LA,  {\tt llamot@lsuhsc.edu}}
\end{center}

\begin{abstract}
SAS introduced Type III methods to address difficulties in dummy-variable models for effects of multiple factors and covariates.  Type III methods are widely used in practice; they are the default method in many statistical computing packages.  Type III sums of squares (SSs) are defined by an algorithm, and an explicit mathematical formulation does not seem to exist.  For that reason, their properties have not been rigorously proven. Some that are widely believed to be true are not always true. An explicit formulation is derived in this paper. It is used as a basis to prove fundamental properties of Type III estimable functions and SSs. It is shown that, in any given setting, Type III effects include all estimable ANOVA effects, and that if all of an ANOVA effect is estimable then the Type III SS tests it exactly. The setting for these results is general, comprising linear models for the mean vector of a response that include arbitrary sets of effects of factors and covariates.
\end{abstract}

\vspace{.5cm}\noindent{\sc Key Words}: ANOVA Effects, Type III Effects, Partitioning SS

\section{Introduction}
Type III estimable functions, hypotheses, and sums of squares came to light in SAS publications in the 1970s, mainly Goodnight (1976) and SAS (1978).  
The recipes given there and in SAS documentation are detailed, but it is difficult to discern a general algorithm or the rationale behind the construction. 

Type III methods are defined in exclusive reference to dummy-variable formulations of multiple linear regression models for factor effects. 
They address fundamental problems encountered in analysis of variance (ANOVA) and known almost since R. A. Fisher first expounded it. In balanced settings, there is practically no disagreement about how main effects and interaction effects should be defined and tested.  In unbalanced settings, though, some crucial properties  no longer hold.  ANOVA sums of squares (SSs) either are not distributed as proportional to chi-squared random variables or they do not test the same hypotheses as in balanced settings.  See Herr (1986) for a historical perspective. 

Type III provided answers in situations where before there had been no consensus and certainly no single ideal answer. It has been criticized in strong words.  Milliken and Johnson (1984, p. 185) say, when there are empty cells, ``... we think that the Type III hypotheses are the worst hypotheses to consider ... because there seems to be no reasonable way to interpret them.''  Venables (2000, p. 12) says, ``I was profoundly disappointed when I saw that S-PLUS 4.5 now provides `Type III' sums of squares as a routine option ... .''  The debate on the merits of Type III methodology continues: see Macnaughton (1998), Langsrud (2003), Hector et al. (2010), and Smith and Cribbie (2014).  

The objective in this paper is to provide a concise mathematical description of Type III SSs and to establish some of their properties.  

\bigskip\noindent\fbox{\begin{minipage}{\textwidth}\hspace{.5cm}Appendix \ref{notation} describes notational conventions, the setting of the linear model, and basic results used here.\end{minipage}}  

\bigskip In short, $\bm{Y}$ (a column vector) follows an $n$-variate normal distribution with mean vector $\bm{\mu} = X\bm{\beta}$ and variance-covariance matrix $\sigma^2\id$. Its realized value is $\bm{y}$. The $n\times k$ model matrix $X$ is fixed and known.   The unknown parameters of its distribution are $\bm{\beta}$, a $k$-vector, and $\sigma^2 > 0$. The model for the mean vector is the set of possibilities for $\bm{\mu}$. It is $\{X\bm{\beta}: \bm{\beta}\in\Re^k\}$, the set of all linear combinations of the columns of $X$. Equivalently, it is $\sp(X)$, the linear subspace of $\Re^n$ \emph{spanned} by the columns of $X$. See \ref{notation}.\ref{span}. It is conventional usage to refer to the model simply as $X\bm{\beta}$.

In the general ANOVA framework, the $n$ subjects are observed under combinations of levels of multiple factors. ``Factor-level combination'' is abbreviated FLC here; FLCs are also called \emph{cells}. The population mean of the response under a FLC is called a \emph{cell mean}. Factor \emph{effects} are differences among the cell means.  The focus of analysis is on factor effects.  Effects of \emph{covariates} on the response are often included in models, too  (see Section \ref{covariates}).

As used here, factor names are A, B, C, and so on, and they appear at $a$, $b$, $c$, etc., levels (all positive), so that there are $a_{\bullet}=abc\cdots$ FLCs possible.  Alternatively, $f$ factors at $a_1, \ldots, a_f$ levels are named F$_1$, $\ldots$, F$_f$, and there are $a_{\bullet} = a_1a_2\cdots a_f$ possible FLCs. A FLC is indexed by $\bm{\ell} = (\ell_1, \ldots, \ell_f)$, with $1 \leq \ell_i \leq a_i$, $i=1, \ldots, f$. There are $n_{\bm{\ell}}$ subjects observed under the $\bm{\ell}$-th FLC.  Cells for which $n_{\bm{\ell}}=0$ are called \emph{empty cells}. \emph{Balanced} models or settings have $n_{\bm{\ell}} = m$ for all FLCs; otherwise the setting is \emph{unbalanced}.

The population mean of the response $Y$ under the $\bm{\ell}$-th FLC is denoted $\eta_{\bm{\ell}}$. The $a_{\bullet}$-vector of these means, in lexicographic order on $\bm{\ell}$, is $\bm{\eta}$.  Denote the average of the entries in $\bm{\eta}$ by $\bar{\eta}_{\bullet}$. See \ref{notation}.\ref{bar dot}.

In balanced models,  terms like  ``A main effects,'' ``B main effects,'' and ``AB interaction effects'' have particular, precise meanings.  They are defined implicitly by the  sets of contrasts on the cell sample means that, when squared and summed, are proportional to the sums of squares (SSs) that R. A. Fisher defined in his exposition of analysis of variance  (see Fisher 1938, p. 240, for example). They will be called \emph{ANOVA effects}. See Section \ref{ANOVA effects}.

In unbalanced models, multiple meanings of these effects have coexisted.   Kutner (1974) lists three definitions of main effects.  Speed et al. (1978) list four ``common ANOVA hypotheses'' that define main effects. Further, names may have different meanings as sets of contrasts, depending on the model in question.  Alone, the term ``A effects'' is ambiguous. 

In common usage, and in syntaxes of statistical computing packages, models are specified by lists of factor effects.  For example, the list (1), A, B specifies a model with an intercept and terms representing A and B main effects.   The list causes sets of columns of $X$ to be formulated in a certain way. No definition of ``effects'' is implied. That same sort of language is used here.  However, modifiers will be used for particular, clearly-defined sets of effects, like ANOVA effects and Type III effects.

\section{Type III in a General Framework}\label{Type III}
Textbooks and articles on the subject give alternative descriptions of Type III estimable functions and SSs, but most of them are incomplete or not entirely correct.  
The best definition is the following from SPSS Statistics $>$ SPSS Statistics 20.0.0 $>$ Help $>$ Statistics Base Option $>$ GLM Univariate Analysis $>$  GLM Model:

\begin{quotation}
\noindent This method calculates the sums of squares of an effect in the design as the sums of squares adjusted for any other effects that do not contain it and orthogonal to any effects (if any) that contain it.
\end{quotation}

Corresponding to this description, consider the columns of $X$ to be partitioned as $X=(X_0, X_1, X_2)$, and consider $\bm{\beta}$ to be partitioned accordingly as $(\bm{\beta}_0\pr, \bm{\beta}_1\pr, \bm{\beta}_2\pr)\pr$.  By rules unnecessary to describe at this point, $X_1$ is defined by the ``effect in the design'' that is the target of interest.    $X_0$ is defined by names of ``effects that do not contain it,'' and $X_2$ by names of  ``any effects (if any) that contain it.''  

While  ``contain'' as used in this definition is well-defined,  it plays no role in this section.  However, in dummy-variable formulations of $X$, terms for any given effect generate a linear subspace that contains (in the set sense) the linear subspaces generated by any other effects that it contains (in the sense meant in the definition).

In this section the partition of $X$ and $\bm{\beta}$ may be considered to be any partition whatsoever.  We leave open the possibilities that $X_0=\bm{0}$ or $X_2=\bm{0}$, that is, that either of these may be a single column of $0$s. If, for example, there would otherwise be no columns in $X_2$, we would insert the $n$-vector $\bm{0}$ as a place-holder.

The estimable linear functions of $\bm{\beta}$ take the form $\bm{h}\pr X\bm{\beta}$, where $\bm{h}$ is an $n$-vector.   The word ``effects'' in the definition refers  to sets of estimable functions. 
Assume that $\bm{h}\in\sp(X)$, that is, that $\bm{h}=\P_X\bm{h}$.  That $\bm{h}$ be ``adjusted for $X_0$'' requires that $X_0\pr\bm{h}=\bm{0}$, or $\bm{h}\in\sp(X_0)^\perp\cap\sp(X)$.    See \ref{notation}.\ref{orth compl}.

Estimable functions of $\bm{\beta}_2$ are those that do not involve $\bm{\beta}_0$ or $\bm{\beta}_1$, that is, $\bm{m}\pr X\bm{\beta}$ with $X_0\pr\bm{m} = \bm{0}$ and $X_1\pr\bm{m}=\bm{0}$.  Equivalently, $\bm{m}\in\sp(X_0, X_1)^\perp\cap\sp(X)$. Let $N_{01}$ be a matrix such that $\sp(N_{01})=\sp(X_0, X_1)^\perp\cap\sp(X)$.  Then estimable functions of $\bm{\beta}_2$ are $\bm{m}\pr X\bm{\beta}$ with 
\[X\pr\bm{m} = \left(\begin{array}{c}0\\ 0\\ X_2\pr\bm{m}\end{array}\right) = 
\left(\begin{array}{c}0\\ 0\\ X_2\pr N_{01}\bm{c}\end{array}\right)
\]
for some $\bm{c}$. That an estimable function $\bm{h}\pr X\bm{\beta}$ be orthogonal to all estimable functions $\bm{m}\pr X\bm{\beta}$ of $\bm{\beta}_2$ requires that
\[ (X\pr\bm{h})\pr (X\pr\bm{m}) = \bm{h}\pr X_2(X_2\pr N_{01}\bm{c}) = 0
\]
for all vectors $\bm{c}$, which requires that $\bm{h}\in \sp(X_2X_2\pr N_{01})^\perp$.

Let $X_{2*} = X_2X_2\pr N_{01}$.  Putting these together, the Type III estimable functions  are  $\{(X\pr\bm{h})\pr \bm{\beta}: \bm{h}\in \bc{S}_{3}\}$,
where 
\begin{equation}\label{S3}
\bc{S}_{3}=\sp(X_0, X_{2*})^\perp\cap\sp(X).
\end{equation}

Let $\P_{3}$ be the orthogonal projection matrix onto the linear subspace $\bc{S}_{3}$, so that $\P_3 = \P_X - \P_{(X_0, X_{2*})}$. 
 Given an $n$-vector of realized values $\bm{y}$, the Type III numerator SS is $SS_3=\bm{y}\pr \P_{3}\bm{y}$, and its df (degrees of freedom) is $\tr(\P_{3})$. Its ncp (non-centrality parameter: see \ref{notation}.\ref{F stat}) is $\bm{\delta}_3\pr\bm{\delta}_3/\sigma^2$, where $\bm{\delta}_3 = \P_3X\bm{\beta}$, and it is 0 iff $\P_3X\bm{\beta}=\bm{0}$.  

$\P_3$ can be computed in several ways. One way is in two steps, with the Gram-Schmidt (GS) construction as described in LaMotte (2014). From GS on $(X_0, X_1, X)$, take $N_{01}$ as the columns in the orthonormal spanning set contributed by $X$ after $(X_0, X_1)$. Compute $X_{2*}$, then compute $Q_3$ as the columns in the orthonormal spanning set from  GS on $(X_0, X_{2*}, X)$ contributed by $X$ after $(X_0, X_{2*})$. Then $\P_3 = Q_3Q_3\pr$, and its df is the number of columns in $Q_3$.

$\bc{S}_3$ defines the set of all Type III estimable functions generated by the target name and its containment relations to the rest of the model.   The direct role of the target name (to which $X_1\bm{\beta}_1$ corresponds)  seems to be peripheral, appearing only through $N_{01}$ in $X_{2*}$. The construction is driven mainly by the rest of the names in the model, those ``adjusted for'' and ``that do not contain'' the target name. It is a Michelangelo construction, trimming away everything else to leave only the object of interest.

Conventionally, we would define the effect of interest as a set of contrasts on the cell means.  The null hypothesis would then be that all of these contrasts are zero. Then we would derive a numerator SS as the restricted model - full model (RMFM) difference in SSE.  

Here, the construction of the test statistic is driven entirely by the definition of Type III estimable functions. The effect in question is not defined directly.  The Type III estimable functions comprise a linear subspace, and the Type III SS  is the squared norm of of the orthogonal projection of $\bm{y}$ onto that subspace. The construction defines a sum of squares. It is not generated by any hypothesis.  

To illuminate $SS_3$ further, recall that the numerator SS can be computed as the  RMFM difference in SSE. 
In $\P_3$, the full model is $\sp(X)$ and the restricted model is $\sp(X_0, X_{2*})$.  It can be shown that $\sp(X) = \sp(X_0, X_1, X_{2*})$ and that $\sp(X_0, X_1, X_{2*})$ is the direct sum  of $\sp(X_0, X_1)$ and $\sp(X_{2*})$.  The restricted model is formed by omitting $X_1$ from the full model.  Because $\sp(X_0, X_1)\cap\sp(X_{2*}) = \{\bm{0}\}$, the estimable functions of $\bm{\beta}_0$ and $\bm{\beta}_1$ are the same in the model $\sp(X_0, X_1)$ and the model $\sp(X_0, X_1, X_{2*})$. See \ref{notation}.\ref{estimable}.

Define $X_{1|0}$ to be $(\id-\P_{X_0})X_1$. It is often called  ``$X_1$ adjusted for $X_0$.''   It can be shown that $\P_{X_{1|0}} = \P_{(X_0, X_1)} - \P_{X_0}$.

The so-called Type II numerator sum of squares is the RMFM SS with $\sp(X_0, X_1)$ as the full model and $\sp(X_0)$ as the restricted model.  In the model $\sp(X_0, X_1)$, it tests exactly (see \ref{notation}.\ref{tests exactly})
\[\text{H}_{02}: (\P_{(X_0, X_1)} - \P_{X_0})(X_0\bm{\beta}_0 + X_1\bm{\beta}_1)  = X_{1|0}\bm{\beta}_1 = \bm{0}.\]
 However, in the full model $(X_0, X_1, X_{2*})$ it tests exactly that $\bm{\delta}_2 = \bm{0}$, where
\begin{eqnarray*}
\bm{\delta}_2 &=& X_{1|0}\bm{\beta}_1 + \P_{X_{1|0}}X_{2*}\bm{\beta}_2.
\end{eqnarray*}

Turning to the Type III SS, working back from the ncp, $SS_3$ tests exactly that $\bm{\delta}_3 = \bm{0}$, that is, 
\[ \text{H}_{03}: (\P_{(X_0, X_1, X_{2*})} - \P_{(X_0, X_{2*})})(X_0\bm{\beta}_0 + X_1\bm{\beta}_1 + X_{2*}\bm{\beta}_{2*})  = \bm{0}.
\]
Verify that
\begin{eqnarray*}
\bm{\delta}_3
 &=& (\P_X-\P_{(X_0, X_{2*})})X_1\bm{\beta}_1\\
&=& (\P_X-\P_{(X_0, X_{2*})})(X_{1|0}\bm{\beta}_1 + \P_{X_0}X_1\bm{\beta}_1)\\
&=& X_{1|0}\bm{\beta}_1 - \P_{(X_0, X_{2*})}X_{1|0}\bm{\beta}_1.
\end{eqnarray*}
Clearly  $\bm{\delta}_3$ is $\bm{0}$ if $X_{1|0}\bm{\beta}_1 = \bm{0}$. Conversely, $\bm{\delta}_3$ can be expressed equivalently as $X_0\bm{\gamma}_0 + X_{1|0}\bm{\beta}_1 + X_{2*}\bm{\gamma}_2$ for some $\bm{\gamma}_0$ and $\bm{\gamma}_2$. The linear subspaces $\sp(X_0)$, $\sp(X_{1|0})$, and $\sp(X_{2*})$ meet only at $\bm{0}$, and so their sum is a direct sum. Then that $\bm{\delta}_3=\bm{0}$ implies that all three components are $\bm{0}$, and in particular that $X_{1|0}\bm{\beta}_1 = \bm{0}$.  This establishes the following proposition.

\begin{propn}\label{H_03}
The Type III SS tests exactly  H$_{03}: X_{1|0}\bm{\beta}_1 = \bm{0}$ in the model $\sp(X_0, X_1, X_{2*})$.\end{propn}

\vspace{.3cm}Note that $X_{1|0}\bm{\beta}_1$ is estimable in the model $\sp(X_0, X_1)$, corresponding to Type II, and in the model $\sp(X_0, X_1, X_{2*})$, corresponding to Type III.  Consequently the dfs for the Type II SS and the Type III SS are the same, namely $\tr(\P_{X_{1|0}})$. They are discussed more fully below.

Now we can see a rationale for transforming $X=(X_0, X_1, X_2)$ to $X_* = (X_0, X_1, X_{2*})$.  If $\sp(X)$ is not the direct sum of $\sp(X_0, X_1)$ and $\sp(X_2)$ (if, e.g., $\sp(X_2)\supset\sp(X_1)$), then H$_{03}$ might not be testable (estimable) in the full model.  It is testable in $\sp(X_0, X_1)$.  Without changing $X_0$ or $X_1$, if we can re-express $\sp(X)$ as a direct sum of $\sp(X_0, X_1)$ and another linear subspace, say $\bc{S}_{2\backslash 01} = \sp(X_{2\backslash 01})$, then H$_{03}$ will be testable in the re-expressed full model. 

The choice of $\bc{S}_{2\backslash 01}$ (and the matrix $X_{2\backslash 01}$ used to generate it) affects the resulting extra SSE.
Suppose, for example, we choose $X_{2\backslash 01}=X_{2|01} = (\id - \P_{(X_0,X_1)})X_2$.  Then $X_0$, $X_{1|0}$, and $X_{2|01}$ are pairwise orthogonal matrices and
\[ \P_X - \P_{(X_0, X_{2|01})} = \P_{X_{1|0}},
\]
and the resulting numerator SS is precisely the Type II SS.

  Another possibility is to define $X_{2\backslash 01}$ to comprise a set of columns of $X_2$ whose span completes $\sp(X_0, X_1)$ to $\sp(X)$ but does not intersect $\sp(X_0,X_1)$ non-trivially.  Each possible choice renders $X_{1|0}\bm{\beta}_1$ estimable in its parameterization of $\sp(X)$, and different choices yield different numerator SSs of the test statistic.  Different choices result in different null spaces in $\sp(X)$; while they may all look like they are testing the same hypothesis, they are actually testing different hypotheses in terms of the mean vector: effects of $X_1$ adjusted for $X_0$ are implicitly defined differently. This may be clearer when it is noted that, in each version of the model, $X_{1|0}\bm{\beta}_1$ is the (generally non-orthogonal) projection of $\bm{\mu} = X\bm{\beta}$ onto $\sp(X_{1|0})$ along $\sp(X_0, X_{2\backslash 01})$.

Among the many possible choices for $X_{2\backslash 01}$, the Type III choice $X_{2*}$ has the attractive orthogonality property that is inherent in balanced ANOVA models. 

\section{Dummy-Variable Models for Factor Effects}

For $f$ factors at $a_1, \ldots, a_f$ levels, a dummy-variable  formulation of a model for the vector $\bm{\eta}$ of cell means is a linear subspace spanned by the columns of a matrix $E$, $\sp(E)$, or described as  $E\bm{\beta}$.  
$E$ is constructed from Kronecker products of identity matrices and vectors of ones. Using a notational scheme that is similar to others used before in this setting, a model is specified by a set  $\bc{J}=\{\bm{j}_1, \ldots, \bm{j}_t\}$. Each $\bm{j}_{i}$ is a binary $f$-tuple that signifies the name of an effect.  That is, $\bm{j}_{i} = j_{i 1}\ldots j_{i f}$, where each ``bit'' $j_{i k}$ is 0 or  1 to indicate absence or presence of the $k$-th factor's token (e.g., A, B, etc.) in the string.  The model in terms of factors A and B signified by the list (1), A, B, AB, for example,  is specified by $\bc{J} = \{ 00,  10,  01,  11\}$, where 00 signifies the absence of both tokens (often denoted (1)); 10 and 01 signify  A and B, respectively; and 11 signifies the name AB. 

Denote the set of all such binary $f$-tuples by $\bc{B}^f$. 
``Containment'' is defined on $\bc{B}^f$. For two such tuples,  $\bm{j}_2$ \emph{contains} $\bm{j}_1$ ($\bm{j}_1$ \emph{is contained in} $\bm{j}_2$) iff $j_{2k} \geq j_{1k}$, $k=1, \ldots, f$. This is denoted $\bm{j}_2\succeq \bm{j}_1$, or as $\bm{j}_2\succ \bm{j}_1$ to exclude $\bm{j}_2 = \bm{j}_1$.  Define $\bm{j}_1\preceq \bm{j}_2$ and $\bm{j}_1 \prec \bm{j}_2$ equivalently.

 For each binary  $\bm{j}\in\bc{B}^f$, $E_{\bm{j}}$ is defined by
\begin{equation}\label{E_j}
E_{\bm{j}} = \bigotimes_{k=1}^f \left\{\begin{array}{l}\bm{1}_{a_{k}} \text{ if } j_{k}=0,\\
\id_{a_{k}}\text{ if } j_{k} = 1.\end{array}\right.
\end{equation}
See \ref{notation}.\ref{Kronecker}.

The $E$ matrix for the model for the cell means in terms of dummy variables is formed by concatenating these matrices for the effects listed in $\bc{J}$: $E_{\bc{J}}=(E_{\bm{j}_1}, \ldots, E_{\bm{j}_t})$.  For $\bc{J}=\{00, 10, 01, 11\}$, 
\begin{equation}E_{\bc{J}}=(E_{00}, E_{10}, E_{01}, E_{11}) = (\bm{1}_a\otimes \bm{1}_b, \id_a\otimes\bm{1}_b, \bm{1}_a\otimes\id_b, \id_a\otimes\id_b).
\end{equation}

\section{ANOVA Effects}\label{ANOVA effects}
\emph{Effects} are differences among the cell means.  Effects are linear functions of $S_{a_{\bullet}}\bm{\eta} = (\eta_{\bm{\ell}} - \bar{\eta}_{\bullet})$.  There are no differences among the cell means iff $S_{a_{\bullet}}\bm{\eta} = \bm{0}$.

For each $\bm{j}\in\bc{B}^f$, define
\begin{equation}\label{Hj}
H_{\bm{j}} = \bigotimes_{k=1}^f 
\left\{ \begin{array}{l} U_{a_{k}}\text{ if } j_{k} = 0,\\
S_{a_{k}}\text{ if } j_{k}=1.\end{array}\right.
\end{equation}
For an $f$-tuple $\bm{j}$ or the name that it signifies, we shall define the  \emph{ANOVA $\bm{j}$ effects} as  $\{\bm{c}\pr\bm{\eta}: \bm{c}\in\sp(H_{\bm{j}})\}$.  For each $\bm{j}$, $\bm{\eta}\pr H_{\bm{j}}\bm{\eta}=\bm{\delta}_{\bm{j}}\pr\bm{\delta}_{\bm{j}}$ times $m/\sigma^2$ is the ncp of the $\bm{j}$-effect ANOVA SS $m\bm{\bar{y}}\pr H_{\bm{j}}\bm{\bar{y}}$ 
in balanced models with $m$ observations per cell.  See \ref{notation}.\ref{F stat}. 
As examples with two factors at $a$ and $b$ levels,  $H_{00}\bm{\eta} = (U_a\otimes U_b)\bm{\eta} = (\bar{\eta}_{\bullet})$, $H_{10}\bm{\eta} = (S_a\otimes U_b)\bm{\eta} = (\bar{\eta}_{i\cdot}-\bar{\eta}_{\bullet})$, $H_{01}\bm{\eta} = (U_a\otimes S_b)\bm{\eta} = (\bar{\eta}_{\cdot j}-\bar{\eta}_{\bullet})$, and $H_{11}\bm{\eta} = (\eta_{ij}- \bar{\eta}_{i\cdot} - \bar{\eta}_{\cdot j} + \bar{\eta}_{\bullet})$. We shall say that a linear function $\bm{c}\pr\bm{\eta}$ is an ANOVA  effect iff $\bm{c}\in\sp(H_{\bm{j}})$ for some $\bm{j}\in\bc{B}^f$. This is a special property, and most linear functions of $\bm{\eta}$ are not ANOVA effects.

These matrices (\ref{Hj}) are symmetric, idempotent, and pairwise orthogonal.  Any sum of distinct $H_{\bm{j}}$ matrices is an orthogonal projection matrix. We shall say that a model (a linear subspace) $\bc{E}$ for $\bm{\eta}$ is a \emph{factor-effects model} iff there is a subset $\bc{J}$ of $\bc{B}^f$ such that $\bc{E} = \sp(H_{\bc{J}})$, with $H_{\bc{J}}  = \sum\{H_{\bm{j}}: \bm{j}\in\bc{J}\}$. 

Let $\bm{j}_0=\bm{j}_{L}0\bm{j}_{R}$ and $\bm{j}_1=\bm{j}_L 1\bm{j}_R$ be two $f$ tuples, the same except that one has 0 and the other has 1 in the $k$-th place. Then
\begin{equation}
H_{\bm{j}_0} + H_{\bm{j}_1} = H_{\bm{j}_L}\otimes (U_{a_k} + S_{a_k})\otimes H_{\bm{j}_R} = H_{\bm{j}_L}\otimes \id_{a_k}\otimes H_{\bm{j}_R}.
\end{equation}
For an $f$-tuple $\bm{j}_*$, it follows that
\begin{equation}\label{pairs}
\sum\{H_{\bm{j}}: \bm{j}\in\bc{B}^f,\;\bm{j}\preceq\bm{j}_*\}  = \bigotimes_{k=1}^f \left\{\begin{array}{l} U_{a_k} \text{ if } j_{*k}=0,\\
\id_{a_k} \text{ if } j_{*k}=1.\end{array}\right.
\end{equation}
Then, with $\bm{j}_*=1\ldots1$,
\begin{equation}\label{sum j*}
\sum\{H_{\bm{j}}: \bm{j}\in\bc{B}^f\} = \id_{a_1}\otimes \cdots \otimes \id_{a_f} = \id_{a_{\bullet}}.
\end{equation}
The same result  (\ref{sum j*}) can be seen by expanding
\[ \id_{a_{\bullet}} = \bigotimes_{k=1}^f \id_{a_k} = \bigotimes_{k=1}^f (U_{a_k} + S_{a_k}).
\]
Noting that $H_{0\ldots0} = U_{a_1}\otimes \cdots \otimes U_{a_f} = U_{a_{\bullet}}$, it follows that
\begin{equation}\label{ANOVA ID}
S_{a_{\bullet}} = \sum\{H_{\bm{j}}: \bm{j} \succ 0\ldots0\}.
\end{equation}
This is the \emph{ANOVA Identity}. It shows that differences $S_{a_{\bullet}}\bm{\eta} = (\eta_{\bm{\ell}}-\bar{\eta}_{\bullet})$ among cell means can be resolved into the sum of $2^f -1$ orthogonal components $H_{\bm{j}}\bm{\eta}$ that are ANOVA effects.  

These relations are needed to establish relations between dummy-variable models and ANOVA effects. Note that, for $\bm{j}_* \in \bc{B}^f$,
\begin{eqnarray}
\P_{E_{\bm{j}_*}}& =& \bigotimes_{k=1}^f \left\{\begin{array}{l} U_{a_{k}}\text{ if }j_{*k}=0,\\
\id_{a_{k}}\text{ if } j_{*k}=1,\end{array}\right.\\
&=& \sum\{H_{\bm{j}}: \bm{j}\in\bc{B}^f \text{ and } \bm{j}\preceq\bm{j}_*\},
\end{eqnarray}
by (\ref{pairs}).  For any subset $\bc{J}$ of $\bc{B}^f$ (that is, for any set of names of effects) and with $E_{\bc{J}}$ concatenating $\{E_{\bm{j}}: \bm{j}\in\bc{J}\}$, $\sp(E_{\bc{J}}) = \sp(\sum\{H_{\bm{j}}: \bm{j}\in\bar{\bc{J}}\})$, where $\bar{\bc{J}}$ is the set of $f$-tuples contained in at least one member of $\bc{J}$. See \ref{notation}.\ref{jbar}.

A dummy-variable model for $\bm{\eta}$ is a factor-effects model.  However, it is interesting to note that this does not hold for full-rank reparameterizations of dummy variable models in which one column is omitted from each $\id_{a_{k}}$ in (\ref{E_j}). (This is  ``reference-level coding.'') As a particular example, the $E$ matrix formed in this way for the model $\{00, 01, 11\}$, which might be considered the restricted model for testing A main effects, is not a factor-effects model.  The extra SSE due to deleting the $E_{10}$ columns from the model does not test any ANOVA effects. In particular, it does not test ANOVA A effects, $H_{10}\bm{\eta}$.

\section{Type III in Dummy-Variable Models}\label{Type III DV}
ANOVA models in terms of dummy variables   take the general form $\sp(X_0, X_1, X_2) = \sp(\mbk E)$, where $E$ is formed by concatenating matrices $E_{\bm{j}}$ column-wise over some set $\bc{J}$ of $f$-tuples, as described in the previous section.   
$\mbk$ has all entries  0, except that there is exactly one 1 in each row. It has $a_{\bullet}$ columns, corresponding to the FLCs. 
Hocking (2013) shows formulations of factor effects models in these terms; his $W$ is $\mbk$.  Empty cells are FLCs for which no subjects are observed; each empty cell results in the column of $\mbk$ corresponding to that FLC being filled with 0s.

The original exposition of Type III (SAS 1978, for example) is in terms of estimable functions of $\bm{\eta}=E\bm{\beta}$ in the model $\bm{\mu} = \mbk (E\bm{\beta})$.  These are functions $\bm{q}\pr E\bm{\beta}$ with $E\pr\bm{q} \in \sp(E\pr \mbk\pr)$. It is clear that $\sp(E\pr \mbk\pr) = \sp(E\pr \mbk_0\pr)$, where $\mbk_0$ is defined as $\mbk$ would be if each positive cell sample size $n_{\bm{\ell}}$ were replaced by 1.  Then $\mbk_0$ has exactly one 1 in each row and at most one 1 in each column.  This establishes a basic property of Type III, that its set of estimable functions depends only on the pattern of empty cells, and it does not depend otherwise on the distribution of cell sample sizes. 

The set of Type III estimable functions for an effect depends also on the list of effects included in the model, of course; depending on it, it can happen that the estimability of some effects is not affected by empty cells.  So-called ``connected'' designs with additive-effects models provide one well-known example.

The model matrix is $X=\mbk E$.  The Type III partition of it as $X=(X_0, X_1, X_2)$ is dictated by the target effect, say $\bm{j}_*\in\bc{J}$, and containment.  Thus $\bc{J}_0 = \{\bm{j}\in\bc{J}: \bm{j}\not\succeq\bm{j}_*\}$, $\bc{J}_1=\{\bm{j}_*\}$, and $\bc{J}_2 = \{\bm{j}\in\bc{J}: \bm{j}\succ\bm{j}_*\}$.  Then $E_k$ is formed by concatenating the columns of $\{E_{\bm{j}}: \bm{j}\in\bc{J}_{k}\}$,  and then $X_{k} = \mbk E_{k}$, $k=0,1,2$.
 
 Given $X=(X_0, X_1, X_2)$, consider $\bm{\beta}$ partitioned correspondingly as $\bm{\beta}=(\bm{\beta}_0\pr, \bm{\beta}_1\pr, \bm{\beta}_2\pr)\pr$.   Let $E_{1|0} = (\id-\P_{E_0})E_1$ and $E_{2*} = E_2E_2\pr\mbk\pr N_{01}$, where as above $\sp(N_{01}) = \sp(X_0, X_1)^\perp$. Note that $\sp(E_0, E_1) = \sp(E_0, E_{1|0})$. Let $E_* = (E_0, E_{1|0}, E_{2*})$.
 Re-express the model $\sp(X)$ as $\sp(\mbk E_*)$. Think of this as $(X_0, X_1, X_{2*})$, but keep in mind that $X_1 = \mbk E_{1|0}$ is not the same as $X_{1|0} = (\id - \P_{X_0})X_1$.

The Type III estimable functions for effect $\bm{j}_*$ are given by (\ref{S3}), which defines $\P_3$ and Type III SS as $SS_{\bm{j}_*} = \bm{y}\pr \P_3\bm{y}$.  The effects that it tests exactly are the linear functions of $\bm{\delta}_3$, and
\begin{eqnarray}  \bm{\delta}_3 &=& \P_3X\bm{\beta}\\
 &=& (\P_{\mbk(E_0, E_{1|0}, E_{2*})} - \P_{\mbk (E_0, E_{2*})})(\mbk E_0\bm{\beta}_0 + \mbk E_{1|0}\bm{\beta}_1 + \mbk E_{2*}\bm{\beta}_{2*})\nonumber\\
 &=& \mbk E_{1|0}\bm{\beta}_1 - \P_{\mbk (E_0, E_{2*})}\mbk E_{1|0}\bm{\beta}_1.\label{E_1|0}
\end{eqnarray}

 \begin{propn}\label{estimable fns}
In the model $\sp(\mbk E_*)$, estimable linear functions of $E_{1|0}\bm{\beta}_1$ are linear functions of $\bm{\delta}_3$.
\end{propn}
{\bf Proof.} Let $R$ be a matrix such that $R\pr E_{1|0}\bm{\beta}_1$ is estimable in the model $\mbk E_*\bm{\beta}$. Then there exists a matrix $L$ such that 
\begin{equation} E_*\pr \mbk\pr L = \left(\begin{array}{c}E_0\pr\\E_{1|0}\pr \\E_{2*}\pr\end{array}\right) \mbk\pr L = \left( \begin{array}{c}0\\E_{1|0}\pr R \\ 0\end{array}\right).\label{estimability conditions}
\end{equation}
It follows that $R\pr E_{1|0}\bm{\beta}_1 = L\pr \bm{\delta}_3$, because 
\[L\pr \mbk E_{1|0} = R\pr E_{1|0} \text{ and } L\pr \mbk (E_0, E_{2*}) = 0.\] \hfill$\Box$

\vspace{.3cm} By Proposition \ref{H_03},  $SS_3$ tests exactly that $X_{1|0}\bm{\beta}_1 = (\mbk E_1 | \mbk E_0)\bm{\beta}_1 = \bm{0}$.  
$SS_3$ tests all the estimable functions  of $E_{1|0}\bm{\beta}_1$, by Proposition \ref{estimable fns}.  If not all linear functions of $E_{1|0}\bm{\beta}_1$ are estimable, then there may be other linear functions of $\bm{\beta}$ that $SS_3$ tests too.
If all of $E_{1|0}\bm{\beta}_1$ is estimable ($R=\id$), then  $\bm{\delta}_3 = \bm{0}$  implies that $E_{1|0}\bm{\beta}_1 = \bm{0}$; and, by (\ref{E_1|0}), $E_{1|0}\bm{\beta}_1 = \bm{0}$ implies that $\bm{\delta}_3=\bm{0}$. 
In that case, $SS_3$ tests exactly that $E_{1|0}\bm{\beta}_1 = \bm{0}$.  

The model for the cell means is $\bm{\eta}=(E_0, E_{1|0}, E_{2*})\bm{\beta}$. To see the connection between $E_{1|0}\bm{\beta}_1$ and ANOVA effects, note that
\begin{eqnarray}
\P_{E_{1|0}} &=& \P_{(E_0, E_1)} - \P_{E_0}\\
&=& \sum\{H_{\bm{j}}: \bm{j}\in\bar{\bc{J}}_0\cup\bar{\bc{J}}_1\} - \sum\{H_{\bm{j}}: \bm{j}\in\bar{\bc{J}}_0\}\\
&=& \sum\{H_{\bm{j}}: \bm{j}\in\bc{J}_*\},
\end{eqnarray}
where $\bc{J}_* = \bar{\bc{J}}_1\backslash\bar{\bc{J}}_0$ is the set of $\bm{j}$s contained in at least one member of $\bc{J}_1$ and not contained in any member of $\bc{J}_0$. See \ref{notation}.\ref{jbar}.
Let  $H_* = \P_{E_{0|1}}$. $H_*\bm{\eta}$ is a sum of ANOVA effects.  

\begin{propn}\label{H_*E_2} $H_* E_0 = 0$, $H_*E_{1|0} = E_{1|0}$, and $H_* E_{2*} = 0$.
\end{propn}

\vspace{.3cm}{\bf Proof.} That $H_*E_0 = 0$ and $H_*E_{1|0}= E_{1|0}$ is clear from the definition of $H_*$. 
Let $\bm{j}_{2}\in\bc{J}_2$ and $\bm{j}\in\bc{J}_*$.  Then $\bm{j}_{2}\succeq \bm{j}_* \succeq \bm{j}$ $\imp$
\begin{eqnarray*} H_{\bm{j}}E_{\bm{j}_{2}}E_{\bm{j}_{2}}\pr &=& \bigotimes_{i=1}^f \left\{\begin{array}{cl}S_{a_i} & j_{i}=1, j_{2i}=1\\
U_{a_i} & j_{i}=0, j_{2i}=1\\
\bm{1}_{a_i}\bm{1}_{a_i}\pr = a_iU_{a_i} & j_{i}=0, j_{2i}=0\end{array}\right.\\
&=&c_{\bm{j},\bm{j}_2}H_{\bm{j}}.
\end{eqnarray*}
Then
\begin{eqnarray*}
H_{\bm{j}}E_2E_2\pr\mbk\pr(\id-\P_{01}) &=& \left(\sum_{\bm{j}_2}c_{\bm{j},\bm{j}_2}\right) H_{\bm{j}}\mbk\pr(\id-\P_{01}).
\end{eqnarray*}
Because $\sp(H_{\bm{j}})\subset\sp(E_1)$, $\sp(\mbk H_{\bm{j}})\subset\sp(\mbk E_1)$, and therefore $N_{01}\pr\mbk H_{\bm{j}} = \bm{0}$ for all $\bm{j}\in\bc{J}_*$.  With $H_* = \sum\{H_{\bm{j}}: \bm{j}\in\bc{J}_*\}$, it follows that $H_*E_2E_2\pr\mbk\pr N_{01} = 0$.\hfill$\Box$

\vspace{.3cm} Now it follows that $E_{1|0}\bm{\beta}_1 = H_*(E_0, E_{1|0}, E_{2*})\bm{\beta} = H_*\bm{\eta}$ in the model for $\bm{\eta}$. If all of $H_*\bm{\eta}$ is estimable, then the Type III SS test of $\bm{j}_*$ effects tests exactly that $H_*\bm{\eta} = \bm{0}$.  Type III $\bm{j}_*$ effects include the estimable part of $H_*\bm{\eta}$. They may include other effects, too.  

It may be true, but it is not proven, that the only ANOVA effects included among Type III $\bm{j}_*$ effects are the estimable part of $H_*\bm{\eta}$. If so, then the other included effects (contrasts), if any, are not ANOVA effects.

It was noted above that Type II and Type III degrees of freedom are the same, say $\nu_2 = \nu_3$.  The Type II full model is $\sp[\mbk(E_0, E_1)]$. It can be represented in two ways, as
\begin{eqnarray*}
\sp[\mbk(E_0, E_1)] &=& \sp(\mbk E_0) \oplus \sp(\mbk E_1 | \mbk E_0), \text{ and as}\\
&=& \sp(\mbk E_0) + \sp(\mbk E_{1|0}),
\end{eqnarray*}
the second because $\sp(E_0, E_1) = \sp(E_0) \oplus \sp(E_{1|0})$, but the direct sum may not carry over.   

Dimensions of these linear subspaces are dfs, which are the same as the ranks of the matrices that generate them.  Let $\nu_{01} = \dim\sp[\mbk(E_0, E_1)]$, $\nu_0 = \dim\sp(\mbk E_0)$, $\nu_2 = \nu_3 = \dim\sp(\mbk E_1 | \mbk E_0)$, and $\nu_{1|0} = \dim\sp(\mbk E_{1|0})$. Let $\nu_* = \dim\sp(E_{1|0})$: it is the \emph{innate} df of the effect.  Note that $\nu_* \geq \nu_{1|0}$. The effect's df is $\nu_*$ if it is entirely estimable. For main effects of factor A at $a$ levels, for example, $\nu_* = a-1$.

Dimensions of direct sums of linear subspaces are the sums of their respective dimensions; and dimensions of sums of linear subspaces are not greater than the sums of their respective dimensions. It follows that
\begin{eqnarray*}
\nu_{01} &=& \nu_0 + \nu_3\\
&\leq& \nu_0 + \nu_{1|0},
\end{eqnarray*}
and hence that $\nu_3 \leq \nu_{1|0} \leq \nu_*$.

Let $\nu_{*0}$ denote the dimension of the estimable part of $E_{1|0}\bm{\beta}_1$. Because the Type III SS tests this estimable part, it follows that $\nu_{*0} \leq \nu_3$.  If all of $E_{1|0}\bm{\beta}_1$ is estimable, then $\nu_{*0} = \nu_*$, which implies also that $\nu_{*0} = \nu_3 = \nu_{1|0} = \nu_*$.  We showed above a stronger result, that in this case Type III SS tests exactly that $H_*\bm{\eta} = E_{1|0}\bm{\beta}_1 = \bm{0}$, which implies that $\nu_3 = \nu_*$.

When testing the same effect in a given setting (characterized by $\mbk$) and model (characterized by $E$), Type II and Type III degrees of freedom are equal. Within the inequalities just shown, practically any relation is possible.  It is not unusual to see, for example, that none of the effect is estimable ($\nu_{*0} = 0$) and that $\nu_3 = \nu_*$. Often too not all of the effect is estimable ($\nu_{*0} < \nu_*$) but $\nu_3 = \nu_*$, that is, the Type III degrees of freedom is the same as if all of the effect were estimable.

Until now, the nominal effect has been a name that led to the partition of the model. $SS_3$ for the nominal A main effect tests all A contrasts, corresponding to $E_{1|0}\bm{\beta}_1$, only if they are all estimable. If they are not, then it tests those that are, plus some more, up to the innate df $a-1$.    It can happen that no A main effect contrasts are estimable, but $SS_3$ still has $a-1$ df.  It is incorrect to say that $SS_3$ tests A main effects when the meaning of A effects is the set of contrasts tested in balanced models. What should we say it tests?  The simple solution is to name whatever $SS_3$ tests ``Type III A effects.''

\section{Illustration}

\begin{table}\begin{center}
\begin{tabular}{|l||r|r|r|}\hline
A$\downarrow$& B$\rightarrow$ 1 & 2 & 3\\\hline\hline
1 & 50.0  & 22.2 & 65.3\\
  &       & 111.7& 53.2\\
  &       &      & 54.2\\\hline
2 & 101.3 & 65.4 & 99.8\\
  &  42.0 &      & 126.8\\
	&  95.5 &      &      \\\hline
3 &  87.3 & 67.0 & 106.2\\
	&  88.6 & 70.2 &      \\
	& 133.2 &      &      \\\hline
\end{tabular}\end{center}
\caption{Data for the example.  Entries are observed responses $y_{ijs}$ on subjects $s=1, \ldots, n_{ij}$ under level $i$ of factor A and $j$ of factor B.}\label{example}
\end{table}

Table \ref{example} lists the observed responses in a setting with factors A and B at  $a = b = 3$ levels each; the cell sample sizes are $(n_{ij})=(1,\; 2,\; 3,\; 3,\; 1,\; 2,\; 3,\; 2,\; 1)\pr$, $n=18$, and $\mbk = \Diag(\bm{1}_{n_{ij}})$. With no empty cells, the 9 columns of $\mbk$ are linearly independent.  For the model defined by the list (1), A, B, and AB,  $\bc{J}=\{00, 10, 01, 11\}$, and $E_{\bc{J}}=(E_{00}, E_{10}, E_{01}, E_{11})$.  The vector of cell means, $\bm{\eta}$, is a $3\times 3 = 9$-vector.  

In balanced models, the A main effects sum of squares tests exactly that there are no A main effects, that is, that all the A marginal means $\bar{\eta}_{i\cdot}$ are equal; equivalently, $H_{10}\bm{\eta}= (\bar{\eta}_{i\cdot} - \bar{\eta}_{\bullet})=\bm{0}$.
For Type III A effects in the model for $\bm{\eta}$ given by $\bc{J}$, $\bm{j}_* = 10$, so $\bc{J}_1 = \{\bm{j}_*\}$,  $\bc{J}_0=\{00, 01\}$, and $\bc{J}_2 = \{11\}$. Then $\bar{\bc{J}}_1 = \{00, 10\}$, $\bar{\bc{J}}_0 = \{00, 01\}$, and  $\bc{J}_* = \bar{\bc{J}}_1\backslash\bar{\bc{J}}_0 = \{10\}$, and hence $H_* = H_{10}$. All of $H_*\bm{\eta}$ is estimable because
\[ E\pr H_* = E\pr \mbk\pr [\mbk(\mbk\pr\mbk)^{-1}H_*].
\]
In this case, $SS_{3A}$ tests exactly that there are no ANOVA A effects.
 Verify that $SS_{3A} =  3286.4603$ and its df is $\nu_{3A} = 2$.

While it has been stated that, in unbalanced models with no empty cells, the Type III SS is the same as Yates's (1934) Method of Weighted Squares of Means (MWSM) SS, and that the MWSM SS tests exactly the ANOVA effect, I have been unable to find a proof of this widely-held belief.  The argument just given, based on Propositions \ref{H_03}, \ref{estimable fns}, and \ref{H_*E_2}, proves that assertion here for $H_* = H_{10}$ in the saturated two-factor model.

In this model, $\bc{J}_*$ contained only the target  $\bm{j}_* = 10$.  Consider now testing Type III AB effects, $\bm{j}_*=11$, in the model defined by $\bc{J}=\{00, 10, 11\}$, so that $\bc{J}_1=\{11\}$, $\bc{J}_0=\{00, 10\}$, and $\bc{J}_2$ is empty.  Then 
\[\bar{\bc{J}}_1\backslash\bar{\bc{J}}_0 =\{00, 10, 01, 11\}\backslash\{00, 10\} = \{01, 11\}.\] Then $H_*= H_{01}+H_{11} = \id_a\otimes S_b$, and $H_*\bm{\eta} = (\eta_{ij}-\bar{\eta}_{i\cdot})$, which are ``B within A'' effects. All of $H_*\bm{\eta}$ is estimable, and so the Type III AB SS tests exactly that there are no B within A effects. In this model, the Type III AB effects are the ANOVA B within A effects, and the Type III SS tests them exactly.

As an intermediate case in which some parts of effects are estimable, but not all, delete the one observation in the 1,1 cell.  Consider again the model $\bc{J}=\{00,10,01,11\}$.     For Type III A effects ($\bm{j}_* = 10$), the contrasts on $\bm{\eta}$ that $SS_{3A}$ tests can be found from $\mbk\pr Q_3$, where $Q_3\pr Q_3=\id$ and $Q_3Q_3\pr = \P_3$.  Only one non-trivial contrast  on the A marginal means is estimable; it is $\bm{c}_{A1}\pr\bm{\eta}=\bar{\eta}_{2\cdot} - \bar{\eta}_{3\cdot}$.  The other Type III A contrast is  $\bm{c}_{A2}\pr\bm{\eta} = 2(\eta_{12}+\eta_{13}) - (\eta_{22}+\eta_{23}+\eta_{32}+\eta_{33})$, which is not an  ANOVA effect. The Type III A effects comprise the $\nu_{3A}=2$-dimensional subspace of linear combinations of these two contrasts. Verify that $SS_{3A} =  2798.1879$. 

Type III gives 2, 2, and 3 df for  A, B, and AB  effects, a total of 7 df for contrasts among the 8 non-empty cell means. (It is reasonable to conjecture that the Type III contrasts generate the model for $\bm{\eta}$ as a direct sum, but that is not established here.)
If we tested exactly the estimable balance-model effects in this case, we would see only 1 df for each main effect and 3 df for AB effects, a total of 5 df of the potential 7 df for contrasts among the 8 cell means.

As a more extreme case, empty the three $i,i$ cells, $i=1,2,3$.  No ANOVA main effects are estimable.    Of the 5 df for differences among the 6 cell means, only 1 df is for an ANOVA effect. 
Type III analyzes the 5 df into 2, 2, and 1 df for Type III A, B, and AB effects. The single estimable AB contrast is an ANOVA AB interaction effect, given by
\begin{eqnarray*}
\bm{c}_{AB}\pr \bm{\eta} &=& \eta_{12} - \eta_{13} - \eta_{21} + \eta_{23}  +\eta_{31} - \eta_{32}\\
&=& [(\eta_{12}-\eta_{13}) -(\eta_{32} - \eta_{33})] - [(\eta_{21} - \eta_{23}) - (\eta_{31} - \eta_{33})].
\end{eqnarray*}
None of the other 4 contrasts is an ANOVA effect. If only ANOVA effects were considered in this setting, then only 1 df of the potential 5 df for differences among cell means would be examined.

\section{Including Covariates}\label{covariates}
Models that include factor effects, covariates, and factor-by-covariate effects can be expressed in the general framework described here.  Consider one covariate $x_1$, with its values in an $n$-vector $\bm{x}_1$.  Define $\mbk_1 = \Diag(\bm{x}_1)\mbk$. Specify the part of the model for the coefficients of  $\bm{x}_1$  in the model for the mean vector $\bm{\mu}$ by a set $\bc{J}_1 = \{\bm{j}_{11}, \ldots, \bm{j}_{1t_1}\}$ of $f$-tuples. The part of the model matrix involving the covariate is then $\mbk_1 E_{\bc{J}_1}$ in terms of dummy variables.

With two factors, A and B, and a single covariate $x_1$, the model that comprises $(1)$, A, B, and AB effects and a linear term in $x_1$ is specified by $\bc{J}=\{00, 10, 01, 11\}$ and $\bc{J}_1 = \{00\}$. To include in addition A by $x_1$ linear effects, $\bc{J}_1 = \{00, 10\}$. The model for the mean vector is then $\sp(\mbk E_{\bc{J}}, \mbk_1 E_{\bc{J}_1})$.  

With $c$ covariates $x_1, \ldots, x_c$, there are $c+1$ sub-models, specified by $\bc{J}, \bc{J}_1, \ldots, \bc{J}_c$. They may be regarded as factor-effects models for intercepts ($\bc{J}$) and coefficients of $x_1, \ldots, x_c$ ($\bc{J}_1, \ldots, \bc{J}_c$).  In this context, denote $\bc{J}$ by $\bc{J}_0$ and $\mbk$ by $\mbk_0$.

The containment relations defined for Type III apply only within sub-models: for any two effects in different sub-models,  neither contains the other, by definition. The purpose of the rest of this section is to establish that the results established in Section \ref{Type III DV}, Propositions \ref{estimable fns} and \ref{H_*E_2}, extend to models that include sub-models for effects of covariates.

A model for the mean vector $\bm{\mu}$ that includes factor effects, effects of $c$ covariates $x_1, \ldots, x_c$, and covariate-by-factor effects can be  formulated generally as $\bm{\mu} = \bm{\mu}_0 + \bm{\mu}_1 + \cdots + \bm{\mu}_c$, where $\bm{\mu}_i \in \sp(\mbk_i E_{\bc{J}_i})$,   and $\mbk_i=\Diag(\bm{x}_i)\mbk_0$, $i=1, \ldots, c$.  Let $\bm{\eta}_i = E_{\bc{J}_i}\bm{\beta}_i$ denote the models for the intercepts ($\bm{\eta}_0$) and coefficients $\bm{\eta}_1, \ldots, \bm{\eta}_c$ of the covariates.  They are specified by lists of effects, $\bc{J}_0, \ldots, \bc{J}_c$.  For an effect $\bm{j}_{i*} \in \bc{J}_i$, the Type III partition $(X_0, X_1, X_2)$ of columns of $X$ has
\begin{eqnarray}
X_1 &=& \mbk_i E_{\bm{j}_{i*}},\\
X_2 &=& \mbk_i E_{\{\bm{j}\in\bc{J}_i: \bm{j} \succ \bm{j}_{i*}\}},\text{ and}\label{contains} \\
X_0 &=& \text{concat}\left(\mbk_{i} E_{\{\bm{j}\in\bc{J}_{i}: \bm{j} \not\succeq \bm{j}_{i*}\}}, \{\mbk_j E_{\bc{J}_j}, j \neq i \}\right),
\end{eqnarray}
where ``concat'' indicates that the matrices in the list are concatenated column-wise.
The form of $X_2$ follows from the definition of containment, which is restricted to the sub-model $\bc{J}_i$.

Proofs of Propositions \ref{estimable fns} and \ref{H_*E_2} can be extended to this setting fairly readily, although comprehensive notation becomes busy.  In Proposition \ref{estimable fns},  (\ref{estimability conditions}) includes additional conditions $E_{\bc{J}_j}\pr\mbk_j\pr L = 0$ for $j=0, \ldots, c, \; j\neq i$. The proof of Proposition \ref{H_*E_2} goes through with little change, except notation, upon substituting $\mbk_i$ for $\mbk$.

With Propositions \ref{estimable fns} and \ref{H_*E_2} established for this general setting, it follows that the Type III SS for $\bm{j}_{i*}$, $SS_{\bm{j}_{i*}}$, tests the estimable part of $H_*\bm{\eta}_i$, where 
\[H_* = \P_{E_{i1|i0}} = \sum\{H_{\bm{j}}: \bm{j} \in \bar{\bc{J}}_{i1}\backslash \bar{\bc{J}}_{i0}\},\]
$\bc{J}_{i1}=\{\bm{j}_{i*}\}$, and $\bc{J}_{i0} = \{\bm{j}\in\bc{J}_i: \bm{j}\not\succeq \bm{j}_{i*}\}$.  The bars indicate tuples $\bm{j} \in \bc{B}^f$ that are contained in at least one member of the set under the bar. See \ref{notation}.\ref{jbar}. 
$SS_{\bm{j}_{i*}}$ tests other contrasts on $\bm{\eta}_{i}$ up to df $=\dim\sp(X_1|X_0) \leq \dim\sp(H_*)$.  If $H_*\bm{\eta}_i$ is estimable, then $SS_{\bm{j}_{i*}}$ tests exactly H$_0: H_*\bm{\eta}_i = \bm{0}$. 

\section{Concluding Comments}
After its introduction, Type III soon became the default method for assessing effects, and it has been regarded with skepticism almost to the point of scorn.  While I used it routinely in analyses, I was also a Type III skeptic.  I think the skepticism was in part because it seemed to be a black box, in part because it was invented by SAS, not published in a rigorously-reviewed and respected statistics journal, and in part due to resentment of what some regarded as SAS's hegemony among statistical computing packages.  Extensive Google searches indicate that this attitude toward Type III is widespread.

If one is determined to test ANOVA main effects of factor A, say, and not all of its df are estimable, then one should identify the estimable part and test it in the context of the general linear hypothesis (see \ref{notation}.\ref{gen lin hyp}), while at the same time stating that the test has no power to test the non-estimable part.

It seems to me that in many applications, perhaps even most, the objective is more exploratory than confirmatory.  The objective is better served then by a systematic look at a comprehensive partition of effects.  Type III partitions the available estimable dfs into up to $2^f-1$ parts, while focusing narrowly only on the  ANOVA effects might produce only a small portion of the dfs available.

Some of the properties of Type III that are widely believed, but not proved before, have been established here in a general framework.  If all of an ANOVA effect $H_*\bm{\eta}$ is estimable, then Type III tests it exactly.  In any case, Type III tests the estimable part of the ANOVA effect, and it tests additional contrasts up to at most the innate df $= \tr(H_*)$ of the target effect.

Other widely-held beliefs are not true in all cases.  Herr (1986) quotes some that are relevant here, in the context of unbalanced two-factor settings, with dummy-variable models parameterized ``with the usual side conditions on the parameters.'' In his definitions of ``four exact methods of analyzing unbalanced, two-way, factorial designs,'' he calls the first STP, or ``STandard Parametric,'' describing it as ``Yates's weighted squares of means; SAS Type III in GLM; SS for rows adjusted for columns and interactions; Searle's $R(\alpha|\mu, \beta, \gamma)$ (side conditions in force).'' Nowhere, as far as I have been able to find, before or since 1986 has it been proven that Type III SS is the same as Yates's MWSM SS, in this two-factor model or any other, nor has it been proven that Type III SS can be had as extra SSE when the estimates of the parameters of the model are subjected to ``the usual side conditions.'' Herr (1986) seems to implicitly assume that there are no empty cells. Otherwise Yates's MWSM SS is not defined, and Type III SS is not equal to it, and Type III SS does not test equality of marginal means, contrary to Herr's (1986)  assertion that it is an ``exact method.'' Herr (1986), like other sources and current SAS documentation, states these assertions, which are partially true and unproven, as facts. Even for this specific model, these properties, when they are true, are  not self-evident, and proving them is not a trivial undertaking.

Some of these beliefs are predicated on there being no empty cells.  In that case, of course, all ANOVA effects are estimable, and, as shown here in a general setting, Type III tests them exactly. With extensive searching, I have not been able to find a previous proof of this property. 

It has been shown here that testing effects exactly depends on estimability, not only on all-filled cells. It is widely asserted that Type III SSs are the same as deleted-variables extra SSEs in dummy-variable models with the ``usual conditions'' imposed on the solutions to the normal equations.  That seems to have been observed, but not proved.   Proving it would require defining ``usual'' conditions in a general setting. The assertion is demonstrably not true when some of an effect is not estimable.  It is asserted that Type III SSs are extra SSEs if contrasts are used to formulate models instead of dummy variables.  That also is not true if some dfs of the target effect are not estimable, and a proof that it is true otherwise does not seem to exist. It has even been asserted that Type III SSs are extra SSEs in what is often called ``reference-level'' coding in which one level of each factor is not included in the dummy variables, so that the model matrix then has full column rank (Milliken and Johnson, 1984, p. 149), which is demonstrably incorrect.

Yates (1934) did not assert that the MWSM SS tested exactly any effect, only that it ``provides an efficient estimate'' of $\sigma^2$ from estimates of marginal means that are weighted averages of cell means.  This was a signal contribution to ANOVA methods, and it continued to be regarded as the gold standard for many decades.  However, it applied only to two-factor models with no empty cells, and it has not been extended to more general settings.  A proof that the MWSM SS tested exactly ANOVA effects did not appear until 1981, when Searle et al. (1981, Appendix B) proved that it is equivalent to a SS that can be shown to test exactly equality of marginal means. 

No other method has been established that has been shown to accomplish the same in a general setting, including those in which parts of ANOVA effects are not estimable.  Type III provides a general approach, far beyond what MWSM provided.  Its basic properties, that it tests the estimable part of ANOVA effects, and if all of an effect is estimable then Type III tests it exactly, have been proven here.  In addition, it has been proven that the Type III df for an effect is the same as the Type II df (that in fact they test the same hypotheses, but in different models), in which the effect is adjusted for all non-containing effects in the model.  It is conjectured, but not proven, that all possible contrasts on (non-empty) cell means are contained in some Type III effect, and that no (non-trivial) contrast is common to any two Type III effects. That is, it is conjectured that Type III provides a comprehensive screening into meaningful components of all estimable contrasts among the cell means.

\begin{appendix}
\section{Notation and Background}\label{notation}
\begin{enumerate}
\item ``If and only if'' is abbreviated iff. 
\item  Any matrix (including vectors) named is assumed to have at least one row and one column.  Other than the implicit assumption that items exist and that row and column dimensions work, any additional properties will be stated. As a particular example, no other property of the model matrix $X$ is assumed.  

\item In algebraic expressions for matrices, assume that row and column dimensions are commensurate with the relations and operations. The matrix sum of two matrices $A$ and $B$ is denoted $A+B$, matrix product by $AB$, transpose by $A\pr$, trace by $\tr(A)$, inverse (if it exists) by $A^{-1}$, generalized inverse by $A^-$, column-wise concatenation by $(A, B)$. 

\item Vectors are column vectors, and they are denoted in boldface. 

\item The set of all $n$-vectors is denoted $\Re^n$.   The linear subspace \emph{spanned} by the $c$ columns of $A$ is $\{A\bm{x}: \bm{x}\in\Re^c\}$ and denoted $\sp(A)$. \label{span}

\item $\id_a$ denotes the $a\times a$ identity matrix, and $\bm{1}_a$ denotes the $a$-vector of ones. The subscripts giving dimensions may be omitted if it is clear in context what they must be.

\item Bar and dot notation:  for subscripted items, overbar indicates averaging and dots in the subscript indicate that the average is over the range of those subscripts.  For example, $\bar{\eta}_{i\cdot} = \sum_{j=1}^b\eta_{ij}/b$, $\bar{y}_{ij\cdot} = \sum_{s=1}^{n_{ij}}y_{ijs}/n_{ij}$, and so on. A subscripted bullet ($\bullet$) indicates that the range is over all subscripts:  $\bar{\eta}_{\bullet} = \sum_{i=1}^a\sum_{j=1}^b \eta_{ij}/ab$.\label{bar dot}

\item ``$(\text{ expression in indices})$'' denotes a vector with entries given by the expression evaluated over the range of indices, in lexicographic order.  With $i=1, \ldots, a$ and $j=1, \ldots, b$, $\bm{\eta} = (\eta_{ij}) = (\eta_{11}, \eta_{12}, \ldots, \eta_{ab})\pr$; $(\bar{\eta_{i\cdot}})$ is the $ab$-vector in which each $\bar{\eta}_{i\cdot}$ is repeated consecutively $b$ times; and $(\bar{\eta}_{\bullet})$ is the $ab$ vector with all its entries equal to $\bar{\eta}_{\bullet}$.\label{ij}

\item $U_a = (1/a)\bm{1}_a\bm{1}_a\pr$ and $S_a = \id_a - U_a$.\label{U and S}

For a vector $\bm{w}$, $U_a\bm{w} = (\bar{w}) = \bar{w}\bm{1}_a$, and $S_a\bm{w} = (w_i-\bar{w})$, where $\bar{w}$ is the average of the $a$ entries in $\bm{w}$.

\item The orthogonal complement of a set $\bc{S}$ of vectors in $\Re^n$, denoted $\bc{S}^\perp$, is the set of vectors in $\Re^n$ orthogonal to all vectors in $\bc{S}$. \label{orth compl}

\item  The orthogonal projection matrix onto a linear subspace $\bc{S}$ of $\Re^n$ is a symmetric, idempotent matrix $P$ such that, for each $\bm{y}\in\Re^n$, $P\bm{y} \in \bc{S}$ and $\bm{y}-P\bm{y} \in \bc{S}^\perp$.  Each linear subspace has exactly one orthogonal projection matrix.  \label{orth projxn}

For a matrix $A$, denote the orthogonal projection matrix onto $\sp(A)$ by $\P_A$. 
A basic property of orthogonal projection is that $\bm{y}\in\sp(A)$ iff $\P_A\bm{y}=\bm{y}$.  

If columns of $Q$ form an orthonormal basis for $\sp(A)$ (i.e., $Q\pr Q = \id$ and $\sp(Q)=\sp(A)$), it can be shown that  $\P_A = QQ\pr$. $Q$ can be obtained directly with the Gram-Schmidt construction applied to the columns of $A$. See LaMotte (2014) for these and other relations. 

Another expression is $\P_A = A(A\pr A)^-A\pr$.

\item The Kronecker product of matrices $A$ and $B$ is $A\otimes B = (a_{ij}B)$, a matrix in which each entry $a_{ij}$ of $A$ is replaced by the matrix $a_{ij}B$.  The main property used here is that $\P_{A\otimes B} = \P_A\otimes \P_B$.  Thus, for example,  $\P_{\id_a\otimes \bm{1}_b} = \P_{\id_a}\otimes \P_{\bm{1}_b} = \id_a\otimes U_b$.  In addition, $A\otimes (B+C) = A\otimes B + A\otimes C$ and $(A+B)\otimes C = A\otimes C + B\otimes C$.\label{Kronecker}

\item The discussion here is in the context of a general linear model for an $n$-vector response variable $\bm{Y}$ (realized value $\bm{y}$) of the form $\bm{Y}\sim\N(X\bm{\beta}, \sigma^2\id)$: that is, $\bm{Y}$ follows a multivariate normal distribution with mean vector $\bm{\mu} = \E(\bm{Y}) = X\bm{\beta}$ and variance-covariance matrix $\var(\bm{Y}) = \sigma^2\id$.  $X$ is a given, fixed $n\times k$ matrix, $\bm{\beta}$ is a $k$-vector parameter ranging over $\Re^k$, and $\sigma^2$ is a positive real-valued parameter.  

A \emph{model} for the mean vector $\bm{\mu}$ is a set of possibilities designated for $\bm{\mu}$.  Each model considered here is a linear subspace of $n$-vectors.  The \emph{full} model is  $\sp(X)$. Other models might take the form $\{X\bm{\beta}: \bm{\beta}\in\Re^k\text{ and } G\pr\bm{\beta} = \bm{0}\}$, where $G$ is a given matrix.  Dealing with non-homogeneous restrictions like $G\pr\bm{\beta} = \bm{c}_0 \neq \bm{0}$ would increase the notational burden, but it would not introduce any new features. 

\item The least-squares estimate of the mean vector in the full model $\sp(X)$ is $\hat{\bm{\mu}} = X\hat{\bm{\beta}} = \P_X\bm{y}$.  Let $C$ be a matrix such that $\sp(C)\subset\sp(X)$ and $XC\pr = \P_X$: for example, $C\pr = (X\pr X)^-X\pr$.  Then $\hat{\bm{\beta}} = C\pr\bm{y}$ is a \emph{least-squares solution}, a function of $\bm{y}$ such that $X\hat{\bm{\beta}} = \P_X\bm{y}$ for all $\bm{y}$. 
Error SS is $SSE = (\bm{y}-X\hat{\bm{\beta}})\pr(\bm{y}-X\hat{\bm{\beta}}) = \bm{y}\pr(\id-\P_X)\bm{y}$. Its degrees of freedom are $\nu_E = \tr(\id-\P_X)$.  If $\nu_E > 0$, Mean Squared Error is $MSE = SSE/\nu_E$.  

\item Inference about the fixed effects $X\bm{\beta}$ is conventionally in terms of  $F$-tests of  linear hypotheses of the form H$_0: G\pr\bm{\beta}=\bm{0}$, where $G$ is a given matrix with $k$ rows.   

A linear function $\bm{g}\pr\bm{\beta}$ of $\bm{\beta}$ (where $\bm{g}$ is a $k$-vector) is said to be \emph{estimable}  iff $\bm{g}\in\sp(X\pr)$. For a good discussion of estimability, equivalent definitions, and its role in linear models, see Seely (1977).  Two equivalent definitions are used here. One is that $\bm{g}\pr\bm{\beta}$ is estimable iff $X\bm{\beta} = \bm{0}$ implies that $\bm{g}\pr\bm{\beta} = 0$ or, equivalently, $\sp(X\pr)^\perp \subset\sp(\bm{g})^\perp$. The other is that imposing $\bm{g}\pr\bm{\beta} = 0$ on the model reduces the model. That is, $\{X\bm{\beta}: \bm{\beta}\in \Re^k \text{ and } \bm{g}\pr\bm{\beta}=0\}$ is a proper subset of the full model $\sp(X)$.

The \emph{estimable part} of $G\pr\bm{\beta}$ is $\{\bm{g}\pr\bm{\beta}: \bm{g} \in\sp(G)\cap\sp(X\pr)\}$.  We shall say that $G\pr\bm{\beta}$ is estimable iff $\bm{g}\pr\bm{\beta}$ is estimable for all $\bm{g}\in\sp(G)$: equivalently, $\sp(G)\subset\sp(X\pr)$.   

\item The sum of two subspaces $\bc{S}_1$ and $\bc{S}_2$ of $\Re^n$ is defined as $\bc{S}_1 + \bc{S}_2 = \{\bm{x}\in\Re^n: \bm{x}=\bm{x}_1 + \bm{x}_2 \text{ for some } \bm{x}_1\in\bc{S}_1 \text{ and } \bm{x}_2\in\bc{S}_2\}$.  

The sum is \emph{direct}, denoted $\bc{S}_1 \oplus \bc{S}_2$, iff for any $\bm{x}$ in the sum, $\bm{x}_1$ and $\bm{x}_2$ are unique.  That the sum is direct is equivalent to $\bc{S}_1 \cap\bc{S}_2 = \{\bm{0}\}$.

\item If columns of $X$ are partitioned as $X=(X_1, X_2)$, and $\bm{\beta}$ correspondingly as $(\bm{\beta}_1\pr, \bm{\beta}_2\pr)\pr$, it can be shown that $X_1\bm{\beta}_1$ is estimable iff $\sp(X) = \sp(X_1) \oplus \sp(X_2)$.  In a direct sum, the estimable functions of $\bm{\beta}_1$ in the model $\sp(X)$ are the same as the estimable functions of $\bm{\beta}_1$ in the model $\sp(X_1)$.\label{estimable}

\item The $F$-statistic has the form $F = (SS/\nu)/MSE$, where the numerator SS is $SS = \bm{y}\pr P\bm{y}$, $P$ is an orthogonal projection matrix, $\sp(P)\subset\sp(X)$, and $\nu=\tr(P)$.  

Under the assumed model, $F$ follows an $F$-distribution with $\nu$ and $\nu_E$ degrees of freedom and noncentrality parameter (ncp) $\lambda^2_P = \bm{\delta}_P\pr\bm{\delta}_P/\sigma^2$, where $\bm{\delta}_P = PX\bm{\beta}$.  \label{F stat}

The ncp is 0, and the distribution is central, iff  $\bm{\delta}=\bm{0}$.  
  
\item   We shall say that a test statistic $F$ or its numerator SS \emph{tests} H$_0: G\pr\bm{\beta}=0$ iff $\bm{\delta}_P =\bm{0}$ implies that $G\pr\bm{\beta}=\bm{0}$: that is, iff $\sp(X\pr P)^\perp \subset \sp(G)^\perp$. Often this will be stated in shorter form as ``$SS$ tests $G\pr\bm{\beta}$.''

We shall say that $F$ or $SS$ \emph{tests exactly} $G\pr\bm{\beta}=\bm{0}$ (or $G\pr\bm{\beta}$)  iff $\sp(X\pr P)^\perp = \sp(G)^\perp$.\label{tests exactly}

\item The numerator  SS for testing H$_0$ can be found as the Restricted Model - Full Model (RMFM) difference in $SSE$.   The full model is $X\bm{\beta}$, and the restricted model is $\{X\bm{\beta}: G\pr\bm{\beta} = \bm{0}\}$.  Let $N$ be a matrix such that $\sp(N) = \sp(G)^\perp = \{\bm{\beta}\in\Re^k: G\pr\bm{\beta}=\bm{0}\}$.  The restricted model is $\sp(XN)$, and the RMFM SS is $SS = \bm{y}\pr (\P_X - \P_{XN})\bm{y}$.  It can be shown that this expression is invariant to the choice of $N$ and that $SS$ tests exactly the estimable part of $G\pr\bm{\beta}$.\label{gen lin hyp}

\item For a subset $\bc{J}$ of $\bc{B}^f$, $\bar{\bc{J}}$ denotes the set of $f$-tuples in $\bc{B}^f$ that are contained in at least one member of $\bc{J}$.\label{jbar}
\end{enumerate}
\end{appendix}

\section{Bibliography}
\begin{itemize}\setlength{\topsep}{0cm}\setlength{\labelsep}{0cm}
\setlength{\itemsep}{0cm}\setlength{\parsep}{0cm}
\setlength{\parskip}{0cm}\setlength{\itemindent}{-1cm}
\item[] Fisher, R. A. (1938).  Statistical Methods for Research Workers, 7th Edition.  Oliver and Boyd, London.
\item[] Goodnight, J. H. (1976). The General Linear Models procedure. Proceedings of the First International SAS User's Group.  SAS Institute Inc., Cary, NC.
\item[] Hector, A., von Felten, S., Schmid, B. (2010).  Analysis of variance with unbalanced data: an update for ecology \& evolution.  Journal of Animal Ecology 79: 308-316.
\item[] Herr, D. G. (1986).  On the history of ANOVA in unbalanced, factorial designs: the first 30 years.  The American Statistician 40: 265-270.
\item[] Hocking, R. R. (2013). Methods and Applications of Linear Models, Third Edition.  John Wiley \& Sons, Inc., Hoboken, New Jersey.
\item[] Kutner, M. H. (1974).  Hypothesis testing in linear models (Eisenhart Model I).  The American Statistician, 28(3): 98-100.
\item[] LaMotte, L. R. (2014). The Gram-Schmidt construction as a basis for linear models.  The American Statistician 68: 52-55.
\item[] Langsrud, \O. (2003).  ANOVA for unbalanced data: Use Type II instead of Type III sums of squares.  Statistics and Computing 13:163-167.
\item[] Macnaughton, D. B. (1998).  Which sums of squares are best in unbalanced analysis of variance? MatStat Research Consulting Inc.
\item[] Milliken, G. A., Johnson, D. E. (1984). Analysis of Messy Data, Volume 1: Designed Experiments. Van Nostrand Reinhold Company, New York.
\item[] SAS Institute Inc. (1978).  SAS Technical Report R-101, Tests of hypotheses in fixed-effects linear models.  Cary, NC.
\item[] Seely, J. (1977). Estimability and Linear Hypotheses. The American Statistician 31(3): 121-123.
\item[] Searle, S. R., Speed, F. M., and Henderson, H. V. (1981). Some computational and model equivalences in analyses of variance of unequal-subclass-numbers data.  The American Statistician 35: 16-33.
\item[] Smith, C. E., Cribbie, R.  (2014).  Factorial ANOVA with unbalanced data: A fresh look at the types of sums of squares.  Journal of Data Science 12: 385-404.
\item[] Speed, F. M., Hocking, R. R., Hackney, O. P. (1978).  Methods of analysis of linear models with unbalanced data. Journal of the American Statistical Association, 73(361): 105-112.
\item[] Venables, W. N. (2000).  Exegeses on linear models.  Paper presented to the S-Plus User's Conference, Washington, DC, 8-9th October, 1998. \verb#https://www.stats.ox.ac.uk/pub/MASS3/Exegeses.pdf#
\item[] Yates, F. (1934).  The analysis of multiple classificatioins with unequal numbers in the different classes.  Journal of the American Statistical Association, 29(185): 51-66.
\end{itemize}

\end{document}